\documentclass[12pt]{article}

\newcommand{\be}{\begin{equation}}
\newcommand{\ee}{\end{equation}}
\newcommand{\bn}{\begin{eqnarray}}
\newcommand{\en}{\end{eqnarray}}
\newcommand{\bns}{\begin{eqnarray*}}
\newcommand{\ens}{\end{eqnarray*}}
\newcommand{\pr}{\prime}

\newcommand{\rref}[1]{(\ref{#1})}
\newcommand{\rs}{Ruijsenaars-Schneider}

\newcommand{\diag}{\mathop{\rm\textstyle{diag}}\nolimits}

\begin{document}

\baselineskip=20pt
\renewcommand{\theequation}{\arabic{section}.\arabic{equation}}

\begin{titlepage}
\begin{flushleft}
{\it Yukawa Institute for Theoretical Physics}
\begin{flushright}
YITP-97-7\\
February 26, 1997\\
hep-th/9702182
\end{flushright}
\end{flushleft}

\ \vspace{1.5cm}

\begin{center}
                        
{\Large \bf    The Ruijsenaars-Schneider Model}\\

\vspace{1.5cm}

{\large \bf    Harry W.\, Braden \\}

\vspace{0.5cm}

{\it           Department of Mathematics and Statistics, 
               University of Edinburgh\\ Edinburgh, UK}
                
{\large         and \\

        \bf        Ryu Sasaki}

\vspace{0.5cm}

{\it           Yukawa Institute for Theoretical Physics, Kyoto 
               University\\ Kyoto 606-01, Japan }

\end{center}

\vspace{1cm}

\begin{abstract}\baselineskip=18pt 
We seek to clarify some of the physical aspects of the 
Ruijsenaars-Schneider models. This important class of models
was presented as a relativistic generalisation of the Calogero-Moser models
but, as we shall argue, this description is misleading. It is far
better to simply view the models as  a one-parameter generalisation of
Calogero-Moser models. By viewing the models as describing certain eigenvalue
motions we can appreciate the generic nature of the models.

\end{abstract}
\end{titlepage}

\section{Introduction}

The Calogero-Moser models \cite{models}\ are completely-integrable,
Hamiltonian systems describing (non-relativistic) particle dynamics with 
pairwise interaction potentials of the form $1/x^2$, $1/\sin^2\!x$, 
$1/\sinh^2\!x$ 
(and in general the Weierstrass $\wp$ functions).
The models are rather generic, which accounts for their importance in
various branches of theoretical physics from solid state physics to 
particle physics \cite{ss, pp, other}:
they appear when describing the eigenvalue motion
of certain matrices \cite{evm}; the pole motions of  the solitons
of various PDE's are described by the model (with possible constraints)
 \cite{poles};
the quantum mechanics \cite{qm} of these models has been connected with the
transmission properties of wires \cite{wires} and Conformal Field Theory
 \cite{cft}.
A rich algebraic structure is being uncovered behind the models \cite{alg}.

The successes of  the Calogero-Moser systems  have
naturally led to an expectation that their ``relativistic'' versions, {\em if 
any}, 
might play similar roles in connection with 
integrable relativistic quantum field theories.
Examples of integrable relativistic quantum field theories include the
sine-Gordon model and affine Toda field theories  (the latter being constructed
from the various affine Lie algebras). 
Thanks to the infinite number of conserved quantities which 
characterises the integrability of  quantum field theories, 
no particle creation and annihilation are allowed  in such theories
and their $N$-particle $S$-matrices
are factorised into a product of ${N(N-1)/2}$ two-particle $S$-matrices.
The expectation that an integrable relativistic field theory 
might equivalently and simply be described in terms of some integrable 
``relativistic'' particle dynamics was speculated by Ruijsenaars in \cite{RS2}
 and appears  more explicitly in \cite{RS1}, where Ruijsenaars and Schneider 
describe the motivation lying behind the discovery of their model. 
Here the model was proposed as a ``relativistic'' (or one-parameter $c$, the
velocity of light) generalisation of Calogero-Moser model.
(The model is variously referred to as the ``relativistic''
Calogero-Moser model or Ruijsenaars-Schneider model. For reasons
we later give, we prefer the latter nomenclature.)

Our aim in the following note is to further explicate these models
and in particular the role of ``relativistic invariance".
The viewpoint described below
is that the Ruijsenaars-Schneider system is an important
and rather generic integrable system, but to describe it as expressing
``relativistic particle dynamics" is quite misleading.
The importance of the Ruijsenaars-Schneider system cannot be underestimated:
it arises as a particular form of eigenvalue motion in much the same way
as the Calogero-Moser model does, and this eigenvalue motion is relevant in
many physical settings. Just as the  Calogero-Moser model is related to 
particular solutions of PDE's, the Ruijsenaars-Schneider model is also
connected with particular soliton solutions of for example the KdV,
mKdV and sine-Gordon equations.
It has been connected with the gauged WZW model \cite{GN}.
A rich algebraic structure is also being uncovered \cite{relag}
for the model and
spin-generalisations \cite{RSspin} of the Ruijsenaars-Schneider model are known,
paralleling\footnote{In this context we note that a Hamiltonian formulation
for these spin-generalisations is still lacking.}
the spin-generalisations \cite{CMspin} of the Calogero-Moser model. The 
solitons of the $a_n$ affine Toda field theories with imaginary
coupling constant have been related \cite{BH} to these
spin-generalisations extending the sine-Gordon  soliton and
Ruijsenaars-Schneider  correspondence  mentioned above \cite{BB}.
But clearly if the same model is related to the relativistically invariant
sine-Gordon equation and also the relativistically
noninvariant KdV equation (and others), the simple notion
of ``relativistic particle dynamics" needs clarification.

The first  difficulty one usually encounters when seeking
to describe ``relativistic particle dynamics" is how any theory with a
single time  can be compatible with causality. Any interaction Hamiltonian 
or Lagrangian depending on the coordinates and momenta of the other particles 
in a single time formulation is by 
definition `action-at-a-distance'. The time evolution of the
positions and momenta is determined by the
positions and momenta of the other particle {\em at the same time}.
In order for this to happen  each particle must be able to `know' the 
coordinates and momenta of the other particles 
{\em instantly}. This  obviously breaks Einstein's causality.
One possible way to circumvent the above difficulty is of course to adopt an
interaction potential of zero range, namely the delta function 
potential.
In two and higher space dimensions  the delta 
function potential is too singular to be treated properly \cite{Kem},
but as is well known in quantum mechanics the delta
function potential in one space dimension can easily be handled.
In fact in this case relativistic many particle theory can be
properly formulated \cite{BNF} and the particle coordinates and times 
obey the Lorentz transformation and together with
the generators of space and time 
translations and boost satisfy the Poincar\'e algebra.
However, with any long range interaction $f(q)$ and a single time formalism
the incompatibility of `action-at-a-distance' with Einstein's
causality remains.
Actually the Ruijsenaars-Schneider models have several ``times"
corresponding to different commuting flows $H_{j}$,
\be
(q(t_1,t_2,\ldots t_l),\theta(t_1,t_2,\ldots t_l))=
\exp\left(\sum_{j=1}\sp{l}t_j H_{j}\right) (q(0),\theta(0)),
\label{evolution}
\ee
and the solutions of the PDE's mentioned above require the evolution
to be determined with respect to each of these times.
In particular, when the flows $H_1$ and $H_{-1}$ are both present
and so $q_j=q_j(t,x)$,
the theory exhibits a Poincar\'e invariance, but as we shall
argue the theory is not relativistically invariant in the sense
suggested by the ``non-relativistic" limit given by Ruijsenaars
and Schneider. Indeed the presence of two ``times" or flows
means we are not dealing with a traditional notion of relativistic
dynamics and the standard ``no-go" theorems \cite{SM}
are correspondingly avoided.
Because the coordinates $q(t,x)$ and $\theta(t,x)$  of the 
Poincar\'e invariant
Ruijsenaars-Schneider model are parameterized by Minkowski space it 
may be thought that what we have here is some, albeit unusual, field
theory. We shall show however that the solutions $q(t,x)$, $\theta(t,x)$ of 
the \rs\ model do not describe the dynamical time-evolution typical of 
field theory and are more akin to those of a topological field theory
in the sense that they do not possess dynamical degrees of freedom.

The Note is organised as follows. In section two some salient features 
of the \rs\ model are briefly reviewed to set the stage and notation. 
We  view the  \rs\ model as describing the motion of eigenvalues
of matrices of certain type, a simple generalisation of the
Calogero-Moser situation.
Then the connection with the $N$-soliton solutions of various soliton
equations (KdV and sine-Gordon, etc) is briefly mentioned.
In section three the nature of the ``relativistic invariance'' of the
\rs\ model is clarified starting with its ``non-relativistic'' limit.
The many ``times'' formulation and the Poincar\'e invariance of the
theory is also discussed.
Section four discusses the field theory aspects of the \rs\ model.
In section five we dwell upon the possible connection between 
integrable quantum field theories with exact factorisable S-matrices and
the \rs\ model. The uncertainty principle of quantum theory plays an
important role here.
The final section is for summary and discussion.
Throughout we will try to use  the  notation of
Ruijsenaars and Schneider \cite{RS1} or Ruijsenaars \cite{RS3}
as far as possible.

\section{The Ruijsenaars-Schneider Model}
\setcounter{equation}{0}
In this section we first review the salient features of the
Ruijsenaars-Schneider model to fix the notation; the details 
may be found in \cite{RS1, RS3}.
Having done this we next review how the model arises when describing
the eigenvalue motion of a particular (possibly partial) differential
matrix equation. This is our perspective on the models, and others may
differ here. Theorems pertaining to these eigenvalue motions may be found in
\cite{RS4}.
We conclude with  the connection between this model and
soliton equations. 

\subsection{Salient Features}
The dynamical variables of the Ruijsenaars-Schneider theory are the
``rapidity'' $\theta_j$ and its canonically conjugate ``position'' 
$q_j$, satisfying the following Poisson bracket relations:
\be
\{q_j,q_k\}=\{\theta_j,\theta_k\}=0,\quad \{q_j,\theta_k\}=\delta_{jk},
\quad j,k=1,\ldots,N.
\label{dynvar}
\ee
We see from (\ref{dynvar}) that if the ``rapidity''  $\theta_j$
is taken to be dimensionless, then $q_j$ has the dimensions of action;
the product of any two canonical variables has the dimensions of action.
The Hamiltonian $H$, the ``space-translation'' generator $P$ and 
``boost'' generator $B$ are given by
\bn
H&=&mc^2\sum_{j=1}^N\cosh\theta_j\prod_{k\neq j}f
\left(\frac{q_j-q_k}{A}\right),\quad
\label{ham}\\
P&=&mc\sum_{j=1}^N\sinh\theta_j\prod_{k\neq j}f
\left(\frac{q_j-q_k}{A}\right),\quad
\label{tran}\\
B&=&-{1\over c}\sum_{j=1}^N q_j,
\label{boost}
\en
where $c$ is the velocity of light and $A$ is a constant having the
dimension of the action (see section three for more detail). 
 They satisfy the following relations
\be
\{H,P\}=0,\quad \{H,B\}=P,\quad \{P,B\}=H/c^2,
\label{Poin}
\ee
provided $f^2(z)$ equals $\lambda+\mu\wp(z)$, including its 
trigonometric, hyperbolic and rational degenerate cases.
These are the relations that the generators of 
the two-dimensional Poincar\'e algebra should satisfy.
It is an added bonus that this choice of the function $f$ also
ensures the existence of $N$ independent, Poisson commuting conserved
quantities, and so the Ruijsenaars-Schneider model is
completely integrable.
Typical  of the conserved quantities constructed are $H_{\pm1}$
where
\begin{equation}
H_{\pm1} = mc^2\sum_{j=1}^N e^{\pm \theta_j}\ \prod^N_{k\not= j}
f\left(\frac{q_j-q_k}{A}\right),
\label{eq:cons}
\end{equation}
and so $H=(H_1 +H_{-1})/2$ and $P=(H_1 -H_{-1})/2$ in the above.

Contrary to \cite{RS1, RS3} we have emphasised the appearance of
the dimensionful parameter $A$ necessary\footnote{ 
In \cite{RS3} Ruijsenaars chooses to work with the
variables $\bar q_j = mc\, q_j$ and $\bar \theta_j= \theta_j/mc$. In this case
a dimensionful length scale $A/mc=L$ must appear in the functions
$(2.22)$ of that reference.}
to define the theory.
The Lagrangians associated with these systems are rather unusual and have
some interesting features. The `Lagrangian' associated with (say) $H_+$ is
\be
{\cal L}=\sum_{j=1}^N \dot q_{j} \left(
\ln\frac{\dot q_{j}}{mc\sp2} -1 -\ln \prod^N_{k\not= j}
f\left(\frac{q_j-q_k}{A}\right)\right),
\label{entlag}
\ee
and we remark that the first 
term on the right here behaves as an \lq entropy\rq.
For the remainder of this section we will set $A=m=c=1$, but will reinstate
these constants at later junctures in our discussion.

\subsection{Eigenvalue Motion and the Ruijsenaars-Schneider model}
The Ruijsenaars-Schneider theory and its generalisations 
may be viewed as describing the
motion of the eigenvalues of matrices of certain type.
For example, let $V$ be a real, symmetric, positive-definite
$N\times N$ matrix whose `time' dependence satisfies
\begin{equation}
\partial V= \Lambda\, V +V\, \Lambda,
\label{eq:def}
\end{equation}
where  $\Lambda$ is a constant matrix.
As we shall now review,
the eigenvalue motion corresponding to (\ref{eq:def})
leads to a mechanical system that is directly
analogous to the linear motions associated with the Calogero-Moser
model. Here the constancy of $\Lambda$ plays the same role as the constants
of motion in the Calogero-Moser situation.
The Ruijsenaars-Schneider model arises when $\partial V$ is further assumed
to be of a specific form; this restriction is directly analogous to the
constraint on the angular momentum made for the Calogero-Moser model.
We will later give examples of such
$N\times N$ matrices satisfying (\ref{eq:def}) that are to be
found in connection with the $N$-soliton solutions of some soliton theories.

Let $V$ be diagonalised by the orthogonal matrix $U$ and set
\[
Q=UVU\sp{-1}=\diag(\exp(q_{1}),...,\exp(q_{N})),
\quad\quad
M=\partial U U\sp{-1},
\]
where $M$ is an anti-symmetric matrix $M=-M^t$.
Then upon setting $L=U\Lambda U\sp{-1}$ we obtain the Lax equation
\be
\partial L=[M,L],
\quad\quad
\partial  Q=[M,Q]+U\partial  V U\sp{-1}=[M,Q]+L\,Q+Q\,L.
\label{laxeq}
\ee
From this it is easy to obtain
\begin{equation}
L_{jj}=(1/2)\partial q_{j} \label{eq:const}
\end{equation}
and (for $j\neq k$)
\[
M_{jk}=\left(\frac{Q_{j}+Q_{k}} {Q_{j}-Q_{k}}
\right)L_{jk}=\coth((q_{j}-q_{k})/2)L_{jk}.
\]
Substituting these into the Lax equation produces 
(with $\dot q_{j}=\partial q_{j}$)
the equations of motion:
\begin{equation}
\dot L_{jj}={1\over2}\ddot q_{j}
=2\sum_{k\neq j}\coth((q_{j}-q_{k})/2)L_{jk}L_{kj},
\label{eq:rsma}
\end{equation}
\begin{equation}   
\begin{array}{l}
\dot L_{jk}=
\frac{1}{2}\coth((q_{j}-q_{k})/2)(\dot q_{k}-\dot q_{j})L_{jk}\\
\quad\quad\quad+\sum_{l\neq j,k} 
(\coth((q_{j}-q_{l})/2)-\coth((q_{l}-q_{k})/2))
L_{jl}L_{lk},\quad (j\neq k).
\end{array}
\label{eq:rsmb} 
\end{equation}  
As shown in \cite{BH} these are 
the spin-generalised Ruijsenaars-Schneider equations \cite{RSspin}\ 
with certain constraints.
 
The (non-spin) model of Ruijsenaars-Schneider  now results when
$\dot{V}$ may be expressed as
\be 
\dot V_{jk}=e_je_k,\quad j,k=1,\ldots, N, 
\label{vdege}
\ee
for some real vector $e$ ($e_j$ being its $j$-th component).
Then with $\tilde{e}=Ue$ we find 
\[ 
L_{jk}=\,\frac{\tilde{e}_{j}\tilde{e}_{k}}{\exp(q_{j})+\exp(q_{k})}. 
\] 
Since we know the diagonal elements of $L$ explicitly in terms of the 
$q_{j}$ we have 
\be 
L_{jk}=\frac{\sqrt{\dot q_{j} \dot q_{k}}}{\cosh((q_{j}-q_{k})/2)}.
\label{laxexp} 
\ee 
This may then be substituted into (\ref{eq:rsma}) to give 
\begin{equation}
\ddot q_{j}=2\sum_{k\neq j}\frac{\dot q_{j}\dot q_{k}}{\sinh(q_{j}-q_{k})}.
\label{eq:rseqm}
\end{equation}
These are the equations of motion for (either $H_{\pm1}$)
\begin{equation}
H_{\pm1} = \sum_{j=1}^N e^{\pm \theta_j}\ \prod^N_{k\not= j}
\coth\biggl({\frac{q_j-q_k}{2} }\biggr) ,
\label{eq:rsham}
\end{equation}
with  conjugate variables $q_j$, $\theta_j$,
satisfying the canonical Poisson bracket relations (\ref{dynvar}).
In this case (\ref{eq:rsmb}) is then identically satisfied.

Now $H=(H_1 +H_{-1})/2$ and $P=(H_1 -H_{-1})/2$ are particular cases of
(\ref{ham}) and (\ref{tran}). Thus the hyperbolic Ruijsenaars-Schneider
model may be identified with the eigenvalue motion just described.
Other (possibly difference \cite{differ}) matrix equations correspond to the
different functions $f$ appearing in the Ruijsenaars-Schneider model.
Further, if $L$ is
the Lax matrix associated with the Ruijsenaars-Schneider theory above then
each of the  flows corresponding to $H_k=(1/k) tr\,L\sp{k}$
is  also conserved and $\{H_k,H_l\}=0$; these give the conserved
quantities associated with the model. Upon setting
\be{\cal H}_k=(H_k+H_{-k})/2,\quad \quad {\cal P}_k=(H_k-H_{-k})/2
,\quad \quad {\cal B}= -\sum_j\sp{N} q_j,
\label{hkb}
\ee
we have
\be
\{{\cal H}_k,{\cal P}_k\}=0,\quad \quad \{{\cal H}_k,{\cal B}\}={\cal P}_k,
\quad \quad \{{\cal P}_k,{\cal B}\}={\cal H}_k.
\label{poink}
\ee
For any $k$ this has the form of the two dimensional Poincar\'e algebra.
Also note from $\sum_j\ddot  q_j=0$ that ${\cal B}$ evolves linearly with 
respect to the $H_1$ flow.
 
\subsection{Connection with $N$-soliton Solutions}
The Ruijsenaars-Schneider theory appears in the study of $N$-soliton
solutions of equations whose tau functions have the  form 
\begin{equation}
\tau=\sum_{\epsilon}\exp
\left(\sum_{j<k}
\epsilon_{j}\epsilon_{k}B_{jk}+
\sum_{j}\epsilon_{j}\zeta_{j}(t,x)\right).
\label{eq:taufn}
\end{equation}
In the above the $\epsilon$ indicates a summation over all possible
combinations of $\epsilon_{j}$ taking the values $0$ or $1$, and the 
indices $j$ and $k$ take values in $\{1,...,N\}$. The
 expression (\ref{eq:taufn}) 
is a rather generic form of the soliton tau function for an integrable PDE,   
the precise nature of $B_{jk}$ and $\zeta_{j}$ depending on the 
particular PDE being considered. It may be viewed as a degeneration
of the theta function solutions of the PDE given via algebraic geometry in
which the $\epsilon_{j}$'s run over all of the integers.
 
Now in appropriate circumstances this tau function can
be written in terms of determinants.
Thus for the Sine-Gordon equation we have
\[
e^{i\beta\phi}=\frac{\det\hspace{0.05in}(1-V)} {\det\hspace{0.05in}(1+V)},
\]
while for the KdV equation
$$\dot u-u u\sp\prime+u\sp{\prime\prime\prime}=0$$
we have $u=-2(\ln\tau)\sp{\prime\prime}$ where
$$\tau={\det\hspace{0.05in}(1+V)}.$$
In both cases the matrix has the form 
\begin{equation}
V_{jk}=\frac{\sqrt{X_{j}X_{k}}}{\mu_{j}+\mu_{k}}, 
\label{eq:vdef}
\end{equation}
where
\begin{equation}
X_{j}=2\, a_{j}\exp\left(\xi_j({ t,x})\right)
\label{eq:Xdef}
\end{equation}
and 
\be
\xi_j(t,x)=\xi_j(0)+\mu_j\sp3\, t-\mu_j\, x,\quad\quad(KdV),
\label{eq:KdVdef}
\ee
\be
\xi_j(t,x)=\xi_j(0)+\mu_j\sp{-1}x_- +\mu_j\, x_+,\quad\quad(SG).
\label{eq:SGdef}
\ee
For the $x$ flow of the KdV equation and either of the  SG flows 
corresponding to the
light cone coordinates $x_\pm$, the matrix equation (\ref{eq:def})
is satisfied and the Ruijsenaars-Schneider
theory \rref{eq:rsham} ensues. For the SG equation $\mu_j$ is related to a 
rapidity.
For other soliton equations that may be expressed in terms of matrices
of the form (\ref{eq:vdef}) and (\ref{eq:Xdef}) the
\lq times\rq\ linear in $\mu_j\sp{\pm1}$ yield the 
Ruijsenaars-Schneider theory (\ref{ham}).
 
\section{Relativistic Invariance}
\setcounter{equation}{0}

We wish now to examine the ``relativistic invariance'' of the 
theories presented by Ruijsenaars and Schneider as \lq{\em a class
of finite-dimensional integrable systems that may be viewed as
relativistic generalizations of the Calogero-Moser systems.\rq}
In the first part of this section we argue  that the Ruijsenaars-Schneider
theory is not
relativistically invariant in the natural variables suggested by
this description. This is why we believe the description of 
Ruijsenaars-Schneider models as \lq relativistic Calogero-Moser models\rq\
is misleading.
Indeed, the ``non-relativistic" limit
of these models requires an explicit scaling of the dimensionful
coupling constant $A$ needed to define these theories, 
and it is unclear why this should be described as a ``non-relativistic" limit.
Rather the relativistic invariance of  the theories, and that
we feel intended by Ruijsenaars and Schneider, is  more subtle.
We shall go on in the latter subsection to investigate this, but note 
that this relativistic invariance
does not yield relativistically invariant particle {\it dynamics}.

At the outset it is instructive to see in what sense the 
Ruijsenaars-Schneider models yield the corresponding Calogero-Moser models
as non-relativistic  limits.
For the sake of both ease and concreteness consider
$$
f\sp2\left(\frac{q_j-q_k}{A}\right) = 1+
\frac{\alpha\sp2}{\sinh\sp2\left(\frac{q_j-q_k}{A}\right)};
$$
similar results hold for the other potentials.
(Here $A$ is to be identified with $2/\mu$ in (4.12) of \cite{RS1}.)
Under the following scalings (which preserve the Poisson bracket relations
for the new variables $\bar q_j$ and $\bar\theta_j$)
\be
\theta_j= \frac{\bar\theta_j}{c},\quad
q_j= c\, \bar q_j,\quad
\alpha=\frac{v}{c},\quad
A=c\, A^\prime
\ee
we find 
\be
H_{nr}=\lim_{c\rightarrow\infty}\left(H-N m c\sp2\right)
=\frac{m}{2}\sum_{j=1}^N{\bar\theta_j}\sp2 +
\sum_{i\ne j}\frac{m {v}\sp2 }
{\sinh\sp2\left(\frac{\bar q_j-\bar q_k}{A^\prime}\right)}.
\ee
Upon using the identification
\be
q_j=x_jmc\cosh\theta_j,\quad p_j=mc\sinh\theta_j,\quad j=1,\ldots,N.
\label{mincoord}
\ee
where now
\be
\{x_j,x_k\}=\{p_j,p_k\}=0,\quad \{x_j,p_k\}=\delta_{jk},
\quad j,k=1,\ldots,N,
\label{xpvar}
\ee
Ruijsenaars and Schneider then  express this as
\be
H_{nr} =\frac{1}{2}\sum_{j=1}^N\frac{p_j\sp2}{m} +
\sum_{i\ne j}\frac{m {v}\sp2 }
{\sinh\sp2\left( (x_j-x_k)/L\right)},
\ee
which is the Hamiltonian of an  appropriate Calogero-Moser model.
(Here we write $A^\prime=mL$, $L$ being a constant having the
dimension of length. The constant $v$ has the dimension of the velocity.)
Similarly they obtain
\bn
P_{nr}&=&m\sum_{j=1}^N \bar\theta_j=\sum_{j=1}^N p_j,\\
B_{nr}&=&-{m}\sum_{j=1}^N x_j.
\en
As Ruijsenaars and Schneider remark, this limit has required
scaling the coupling constants of the theory.
Indeed, however one takes this limit, one cannot avoid\footnote{
It may at first appear that the $\beta$ scaling given in \cite{RS1}
avoids the scaling of the parameter $\mu$, which plays the role of
$1/A$ here. This is not really the case, for the $q_j$ variables
must also be scaled to preserve the Poisson bracket relations; the
three different scalings given in \cite{RS1, RS3} are identical.}
scaling the dimensionful \lq coupling constant\rq\ $A$.
Certainly this analysis shows that the Ruijsenaars-Schneider models
reduce to the Calogero-Moser models in a particular scaling limit,
but it is not clear that this should be described physically as a
``non-relativistic" limit. Only by (infinitely) shifting the
Hamiltonian do the  generators of the Poincar\'e algebra reduce to the
Galilei generators and, as we shall now show, the Ruijsenaars-Schneider model
is not relativistically invariant in the naive sense one would expect
for a theory described as a ``relativistic generalisation" of the
Calogero-Moser model. It seems altogether better to describe the 
Ruijsenaars-Schneider theory as a one-parameter extension of the
Calogero-Moser models.

Now Einstein's special relativity simply states that an `event' is a point
in {\em Minkowski space}.
The essential point is that special relativity  is more than 
a  closed Poincar\'e algebra (like \rref{Poin}): one also needs the
Minkowski space upon which it acts via the inhomogeneous Lorentz  (Poincar\'e)
transformation
\be
\pmatrix{t^\pr_0\cr x^\pr_0\cr}=\pmatrix{\cosh\alpha &\sinh\alpha\cr
\sinh\alpha &\cosh\alpha\cr}
\pmatrix{t_0\cr x_0\cr}+ \pmatrix{a\cr b\cr}.
\label{pointr}
\ee
For relativistically invariant particle {\em 
dynamics} one further needs {\em dynamical variables} directly related with 
the {\em Minkowski} positions and momenta.
Now by describing their models as \lq relativistic generalisations\rq\
of the Calogero-Moser system, one is naturally led to expect
that the $q_j$ or the $x_j$, arising in the ``non-relativistic limit" above, 
are possible Minkowski space variables.
Indeed if we wish the Hamiltonian (\ref{ham}) to be
space translation invariant --it is manifestly time-translation invariant
since the Hamiltonian $H$ does not contain the time explicitly--
we must identify the $q_j$ as the Minkowski space variables
since (\ref{ham}) depends only on their differences.  Let us see that neither
 $q_j$ or $x_j$ are possible Minkowski space variables.
To this end we record the following actions 
of the ``space-translation'' generator $P$ and ``boost" generator $B$
on $q_j$ and $\theta_j$:
\be
\delta_Pq_j=\{q_j,P\}=mc\cosh\theta_j\prod_{k\neq j}f(q_j-q_k), 
\quad \delta_P\theta_j=\{\theta_j,P\}\neq0,
\label{sptract}
\ee
\be
\delta_B\theta_j=\{\theta_j,B\}={1\over c},\quad\quad 
\delta_Bq_j=\{q_j,B\}=0.
\label{boos}
\ee 
These imply
\be
\delta_Px_j=\prod_{k\neq j}f(q_j-q_k)-x_j\tanh\theta_j\{\theta_j,P\},
\quad \delta_Pp_j\neq0,
\label{sptrxact}
\ee
and that the finite transformations under ``boosts" are
\be
\theta^\pr_j=\theta_j+{\alpha\over c},\quad q_j^\pr=q_j,\quad {\rm or}
\quad 
x^\pr_j=x_j{\cosh\theta_j\over{\cosh(\theta_j+{\alpha\over c})}}.
\label{finboos}
\ee

Now we see from \rref{sptract} and \rref{sptrxact} that neither
$q_j$ nor $x_j$ transform as the coordinates of
the Minkowski space under a space translation --in fact they are changed by
amounts depending on the particle positions and momenta.
Further, although the  rapidities have the correct transformation 
\rref{finboos} that of the
Minkowski positions is very different from the ordinary Lorentz boost.
We conclude therefore that the theory is not relativistically invariant
in the naive sense suggested by the ``non-relativistic" limit
given by Ruijsenaars and Schneider.
Of course the details of the above verification for the non-invariance
under the inhomogeneous Lorentz transformation have depended on our
identification of the Minkowski coordinates and momenta, but without
giving these explicitly the Ruijsenaars-Schneider theory cannot be said to
describe relativistic dynamics.

We have argued  that the Hamiltonian dynamics of the
Ruijsenaars-Schneider theory is not invariant under 
Einstein's special theory of relativity in the naive sense
suggested by the ``non-relativistic" limit
given by Ruijsenaars and Schneider. As such we believe the
description of these models as ``relativistic Calogero-Moser"
systems is thoroughly misleading. Indeed, the ``non-relativistic" limit
of these models requires an explicit scaling of the dimensionful
coupling constant $A$ above, and it is unclear why this should be
described as a ``non-relativistic" limit at all. It seems far more
sensible to view the models as one-parameter generalisations of
the Calogero-Moser systems.

\subsection{Many ``times'' and Poincar\'e Invariance}

It remains to explain in what sense Ruijsenaars-Schneider theory
evidences Poincar\'e invariance. For such an invariance we
require several ``times' and their corresponding flows $H_k$. These
times will be our coordinates. Now the dynamical variables evolve
according to (\ref{evolution})
\be
(q(t_1,t_2,\ldots t_l),\theta(t_1,t_2,\ldots t_l))=
\exp\left(\sum_{j=1}\sp{l}t_j H_{j}\right) (q(0),\theta(0)),
\ee
and because we have several times we are not really dealing with
dynamics. Thus using our description 
(\ref{eq:vdef},\ref{eq:Xdef},\ref{eq:KdVdef})
we see the solitons of the KdV equation evolve according to
$H_1$ and $H_3$ and we have $q_j=q_j(t_1,t_3)$. Similarly the solitons
of the SG  equation evolve according to $H_1$ and $H_{-1}$ and we have
$q_j=q_j(t_{-1},t_1)$. In the further restricted setting when we
are dealing with flows $H_k$ and $H_{-k}$ it is possible to consider the
associated Poincar\'e algebra (\ref{poink}). This is what distinguishes
between the various soliton equations: although we may associate the
Ruijsenaars-Schneider Hamiltonian (\ref{ham})
with solitons of each of the KdV, mKdV and SG equations for example,
only the SG equation has a second flow that yields an associated
Poincar\'e algebra. It remains to check that the ``boost" does indeed behave
correctly. Of course we always have that
$$
e\sp{\alpha {\cal B}}
e\sp{ t_k {\cal H}_k -t_{-k} {\cal P}_k}
e\sp{-\alpha {\cal B}}
=e\sp{ t_k\sp\prime {\cal H}_k -t_{-k}\sp\prime {\cal P}_k},
$$
where
$$
\pmatrix{t^\pr_k\cr t^\pr_{-k}\cr}=\pmatrix{\cosh\alpha &\sinh\alpha\cr
\sinh\alpha &\cosh\alpha\cr}
\pmatrix{t_k\cr t_{-k}\cr},
$$
but when we further have that $e\sp{\alpha {\cal B}}q_j(0)=q_j(0)$
(i.e. when $q_j$ behaves as a Lorentz scalar) we see that
\be
e\sp{\alpha {\cal B}}q_j(t_k,t_{-k})=q_j(t^\pr_k,t^\pr_{-k})
\label{nponik}
\ee
and we have an action of the Poincar\'e algebra on our coordinates.
Using (\ref{finboos}) we see for example that this is true for
the $H_{\pm 1}$ flows for the SG equation. It is in this
sense that the Ruijsenaars-Schneider theory is said to evidence
Poincar\'e invariance, but this is very different from
relativistic particle dynamics. 

Let us  further consider the SG example where $(t_1,t_{-1})=(t,x)$.
Here we have
\be
[q(t,x), \theta(t,x)]_j=[\exp(tH-xP)(q(0),\theta(0)]_j,\quad j=1,\ldots,N,
\label{rseq}
\ee
in which $H$ and $P$ are given by (\ref{ham}) and (\ref{tran}), respectively.
In this very specific setting,
because the $q_j$ behave as Lorentz scalar scalars, we may
define a ``trajectory" via
$$
q_j( t, x_j(t))=0.
$$
Ruijsenaars and Schneider show that this specifies $x_j(t)$ for all time
and (\ref{nponik}) shows these trajectories are Lorentz invariant.
(Indeed we could have set $q_j$ to equal any constant with a similar
result; the choice $q_j=0$ is motivated by the fact that asymptotically
these correspond to the peaks of the solitons.)
Now although these ``trajectories'' are relativistically invariant
we again emphasise that they have not been presented as
relativistically invariant {\em dynamics}.

Before closing this section a further comment is in order. 
We have seen that the two ``times'', $t_1$ and $t_{-1}$, or $t$ and $x$, 
are necessary for the Poincar\'e invariance of the \rs\ model. 
So, logically  both of these times should be
carried
over to its non-relativistic limit, the Calogero-Moser models.
Under the  $x$ evolution we simply have
$$
x_j(t,x)=x_j(t)+x,\quad\quad p_j(t,x)=p_j(t).$$
In the physical interpretation of the Calogero-Moser models
the presence of this additional ``time'' $x$ is both redundant and rather
confusing.
This is another reason why we believe that the description of the \rs\
model as ``relativistic Calogero-Moser models'' is misleading.

\section{Ruijsenaars-Schneider Theory and Field Theory}
\setcounter{equation}{0}
 
In the previous section we have seen that when one considers
the $H_{\pm1}$ (or equivalently the $H$ and $P$) flows associated
with for example the SG equation, the  $q_j(t,x)$ given by
(\ref{rseq}) behave as Lorentz scalars. One might  naively be 
tempted to think these describe an $N$-component scalar field in  the
$1+1$ dimensional Minkowski space $(t,x)$, and similarly
that $\theta(t,x)$ are {\em dynamical} fields of some $1+1$ 
dimensional theory. We will now argue that this is not really the case
and discuss the physical content of the solutions $q(t,x)$,
$\theta(t,x)$ of the \rs\ model.

An ordinary field variable, say $\phi(t,x)$ describes a dynamical
system with infinitely many degrees of freedom (one associated to each 
point $ x$ of space). At equal times these degrees of freedom are independent
of each other and this is expressed by the
Poisson bracket (or commutation) relation 
$$
\{\phi(t,x),\phi(t,y)\}=0,\quad ([\phi(t,x),\phi(t,y)]=0).
$$
In other words, in an initial value problem ($t=0$), the initial values
$\phi(0,x)$ can be chosen arbitrarily.
On the other hand, as is clear from (\ref{rseq}) $q(0,x)$ and  $\theta(0,x)$
are severely constrained. They are the solutions of
\bn
{\partial\over{\partial x}}q_j(0,x)&=&\{q_j(0,x),P\},\nonumber\\
{\partial\over{\partial x}}\theta_j(0,x)&=&\{\theta_j(0,x),P\}, \quad t=0,\quad 
-\infty<x<\infty,\label{constreq}
\en
with the condition $q_j(0,0)=q_j(0)$ and $\theta_j(0,0)=\theta_j(0)$.
It is obvious that such constraints can never be imposed on a relativistic field
variable $\phi(0,x)$ without breaking causality.
 
Further, the ``time-evolution" of  $q_j(t,x)$ and
$\theta_j(t,x)$ are also very different from those of a relativistic field.
At time $t$, $q_j(t,x)$ and
$\theta_j(t,x)$ are solely determined by the ``initial data" $\{q_k(0,x),\theta_
k(0,x)\}$, $k=1,\ldots,N$ depending only on the {\em same} $x$, 
since they are solutions of
\bn
{\partial\over{\partial t}}q_j(t,x)&=&\{q_j(t,x),H\},\nonumber\\
{\partial\over{\partial t}}\theta_j(t,x)&=&\{\theta_j(t,x),H\},\label{timeveq}
\en
with the initial value $q_j(0,x)$ and $\theta_j(0,x)$. This is in 
marked  contrast with a dynamical relativistic field, 
in which $\phi(t,x)$ depends on the initial data
$\phi(0,y)$ within the past light-cone, ie, $x-ct\leq y\leq x+ct$.
 
Indeed, given the $2N$ initial conditions 
$q_j(0,0)=q_j(0)$ and $\theta_j(0,0)=\theta_j(0)$ at any one point,
the solutions $q_j(t,x)$, $j=1,\ldots,N$ of \rs\ models are then 
specified ``globally". The properties we have just described show that
the the solutions $q(t,x)$, $\theta(t,x)$ of the \rs\ model
are not describing the dynamical time-evolution typical of 
field theory. Indeed this lack of ``dynamics" bears many of the 
hallmarks of a topological field theory:
although we cannot as yet make this precise we conclude the section
with a Lax pair encoding the evolution with respect to the
various flows.
{\em
Let $V$ be an $N\times N$ diagonalisable matrix  such that
\begin{equation}
\partial_\pm V= \Lambda_\pm\, V +V\, \Lambda_\pm\quad\quad
[\Lambda_+,\Lambda_-]=0,
\end{equation}
and $\Lambda_\pm$ are constant.
Then with $Q=UVU\sp{-1}=\diag(\exp(q_{1}),...,\exp(q_{N}))$,
$M_\pm=\partial_\pm U U\sp{-1}$ and $L_\pm=U\Lambda_\pm U\sp{-1}$
we have
\begin{equation}
[D_+,D_-]=0
\end{equation}
where
\be
D_\pm =\partial_\pm+\left(
\begin{array}{cc}
-M_\pm-L_\pm&\mu\sp{\pm1} Q\\
0&L_\pm-M_\pm
\end{array}
\right),
\label{dplusmi}
\ee
and  $\mu$ is a spectral parameter.
}

\section{Uncertainty Principle}
\setcounter{equation}{0}
Let us now examine the possibility of `reducing' an integrable
relativistic quantum field theory with factorisable $S$-matrices
to a collection of fixed particle number quantum mechanical systems;
this was mentioned as means of motivation at the outset of the
work of Ruijsenaars and Schneider.
The known exact factorisable $S$-matrices of, for example, 
sine-Gordon theory \cite{Zam}\ 
and affine Toda field theory \cite{BCDSa, CM} have been obtained as
solutions of the Yang-Baxter equation and/or bootstrap equation satisfying 
analyticity, unitarity and crossing symmetry.
Now in a crossing symmetric quantum field theory a field operator 
$\phi_j$ annihilates particles of species $j$ and creates their 
anti-particles. Therefore any interaction term in the Lagrangian
of a crossing symmetric field theory changes the particle numbers.
This is in sharp contrast with  non-relativistic quantum field 
theories (for example, the non-linear Schr\"odinger theory) in which
the interaction term $(\bar\psi\psi)^2$ is manifestly particle number 
preserving:
$\psi$ annihilates a particle and $\bar\psi$ creates a particle.

\bigskip
In contrast to the no-particle production
which is the hallmark of an integrable classical field theory and based on 
its infinite number of conservation laws, 
in relativistic quantum field theory this
property is only guaranteed 
between the two asymptotic states at $t=\infty$
and $t=-\infty$ \cite{Lan}.
In other words, the results of  measuring any classically conserved quantity
over a  finite time interval will fluctuate because of the
uncertainty principle of the quantum theory.
In particular, the particle numbers will not be constant over time 
due to the various virtual processes caused by the above mentioned 
particle number non-preserving interactions.
Various field theoretical calculations of the S-matrices and other 
quantities \cite{BCDSb}\  in affine Toda field theory show this fact 
explicitly.
Thus we arrive at the conclusion that a `reduction' of  a solvable
relativistic quantum field theory to a collection of 
fixed particle number (relativistic) quantum mechanical systems is impossible.
 
\section{Summary and Discussion}
\setcounter{equation}{0}
We have discussed various aspects of the \rs\ model. 
In particular we have argued that these models are most naturally
viewed as a one-parameter generalisation of the Calogero-Moser
models which should not be described as a ``relativistic"
generalisation: the model is not in fact ``relativistically invariant"
in the sense dictated by the ``non-relativistic'' limit.
 Further we have compared the
many (compatible) ``times'' formulation --in which certain models are
Poincar\'e invariant-- with standard field theory.
In this context the \rs\ model does not describe a dynamical field
theory.  This is entirely natural in the soliton setting that gives rise
to the model, for the \rs\ equations simply describe a {\it single}
solution to the associated soliton-bearing PDE in an analogous manner to the
inverse scattering transform.
We have also discussed  constraints that the uncertainty principle
places on any possible linkage
between integrable quantum field theories with  exact factorisable
S-matrices and integrable particle dynamics.
In spite of the difficulties related to the ``relativistic''
interpretation, we again emphasise the importance of these models,
an importance we believe stems from the natural matrix equations
associated with the models. The \rs\ equations in this setting are not only
generic but useful. The work on \rs\ models with
spin degrees of freedom  is still in its infancy and we would
like the connections between such models and the affine Toda field theories
to be pursued both algebraically and physically.

\section*{Acknowledgments}
We thank L.~O'Raifeartaigh and Y.~Munakata for useful discussion.
This work is partially funded by the Royal Society and JSPS (Japan society for 
Promotion of Science), the Anglo-Japanese joint research project.
We thank JSPS and RS for financial support.
H.W.B. thanks YITP and RIMS, Kyoto University for hospitality.
R.S. thanks Dept. Math. Univ. Durham for hospitality.

\end{document}